\documentclass{iaus}
\usepackage{graphicx}

  \checkfont{eurm10}
  \iffontfound
    \IfFileExists{upmath.sty}
      {\typeout{^^JFound AMS Euler Roman fonts on the system,
                   using the 'upmath' package.^^J}%
       \usepackage{upmath}}
      {\typeout{^^JFound AMS Euler Roman fonts on the system, but you
                   dont seem to have the}%
       \typeout{'upmath' package installed. iaus.cls can take advantage
                 of these fonts,^^Jif you use 'upmath' package.^^J}%
      }
  \else
  \fi

  \checkfont{msam10}
  \iffontfound
    \IfFileExists{amssymb.sty}
      {\typeout{^^JFound AMS Symbol fonts on the system, using the
                'amssymb' package.^^J}%
       \usepackage{amssymb}%

      }{}
  \fi

  \IfFileExists{amsbsy.sty}
    {\typeout{^^JFound the 'amsbsy' package on the system, using it.^^J}%
     \usepackage{amsbsy}}
    {}

\newsavebox{\astrutbox}
\sbox{\astrutbox}{\rule[-5pt]{0pt}{20pt}}

\title[Outskirts of Galaxy Clusters: intense life in the suburbs]
      {How does clustering depend on environment?}

\author[Ummi Abbas, Ravi K. Sheth]%
{Ummi Abbas$^1$%
Ravi K. Sheth$^1$}

\affiliation{$^1$Department of Physics \& Astronomy, University of Pittsburgh,
Pittsburgh, PA 15260, USA email: ummi@phyast.pitt.edu}

\pubyear{2004}
\volume{195}
\pagerange{1--8}
\date{?? and in revised form ??}
\jname{Outskirts of Galaxy Clusters: intense life in the suburbs}
\editors{A. Diaferio, ed.}
\begin{document}

\maketitle

\begin{abstract}
According to the current paradigm, galaxies form and reside in
extended cold dark matter (CDM) halos and in turn are key tracers
of cosmological structure. Understanding how different types of
galaxies occupy halos of different masses is one of the major
challenges facing extragalactic astrophysics. The observed galaxy
properties depend on the environment surrounding the galaxy.
Within the framework of most galaxy formation models the
environmental dependence of the galaxy population is mainly due to
the change of the halo mass function with large-scale environment.
Such models make precise predictions for how galaxy clustering
should depend on environment. We will illustrate this by
presenting analytical models of dark matter and galaxy clustering
along with results obtained from numerical simulations. With these
results we can hope to obtain a better understanding of the link
between galaxies and dark matter and thereby constrain galaxy
formation models.
\end{abstract}

\section{Introduction}
The properties of galaxies are determined by the dark matter haloes 
in which they form. The link between galaxies and haloes is an 
imprint of various complicated physical processes related to galaxy 
formation such as gas cooling, star formation, merging,
tidal stripping and heating, and a variety of feedback processes.
Different types of galaxies have different types of evolutionary 
histories and clustering properties. For example, observationally 
it has been found that clustering of galaxies is dependent  on their 
luminosity and color. More luminous galaxies are generally more strongly
clustered than fainter ones, whereas red galaxies are more clustered 
than blue galaxies (Zehavi et al. 2002).
On the other hand, haloes of different mass are clustered differently, 
which in turn are populated by different galaxy types.
We can therefore infer the link between galaxies and dark matter haloes 
directly from the observed clustering properties of galaxies. 
The halo based description is the backbone of this approach and provides 
a simple and natural way of modelling the difference between the 
clustering of galaxies relative to 
dark matter in the form of the halo model (Cooray \& Sheth 2002).

\section{Quantifying Clustering with the Correlation Function}

\subsection*{Analytical modelling}

Within the framework of the halo model the correlation function
can be expressed in terms of two components: 
\begin{equation}
    \xi(r|\delta) = \xi_{1h}(r|\delta)+ \xi_{2h}(r|\delta)
\end{equation}
where $\delta$ is the overdensity in a region of mass M and volume V 
which is $(M/\bar{\rho} V)-1$, $\bar{\rho}$ being the background
density.
The first term is the one halo term, which is the
contribution from pairs of particles inside the same halo, and the
second term is from pairs of particles lying in two different
haloes.

The one halo term depends on the conditional mass function,
$n(m|\delta)$, and on the density profile which is assumed to be
NFW (Navarro et al. 1997, Sheth \& Tormen 2002). It dominates on
small scales and is written as:

\begin{equation}
    \xi_{1h}(r|\delta)= \int_0^{\infty}dm\,n(m|\delta)\frac{m^2}
{p(\delta)\rho^2}\lambda(r|m)
\end{equation}
where $p(\delta)$ is the probability of finding a region of overdensity
$\delta$, $\rho$ is the background density of the overdense regions, and
$\lambda(r|m)$ is the convolution of the density profile with itself 
(Sheth et al. 2001).

The two halo term depends on the conditional mass function and on
the halo-halo correlation function $\xi_{hh}$. 
It dominates on large scales
where linear theory is accurate and can be written as:

\begin{equation}
    \xi_{2h}(r|\delta) =  \int_0^{\infty}dm_1 n(m_1|\delta)\frac{m_1}{\rho}
    \int_0^{\infty}dm_2\,n(m_2|\delta)\frac{m_2}{\rho}\xi_{hh}(r|m_1,m_2)
\end{equation}

The halo model has
become more and more accurate owing to the fact that detailed
analytical descriptions for the structure and clustering of dark
matter haloes have become available.

The halo model can be extended to address the bias of galaxies by
introducing a model for the halo occupation number, $N_g(m)$,
which describes how many galaxies on average occupy a halo of mass
$m$. It is seen that
the environmental dependence of the galaxy population is mainly
due to the change of the halo mass function with large-scale
environment. So we can hope to understand the environmental
dependence of the galaxy population by understanding how galaxies
occupy dark matter haloes of different masses.

\section{Results and Future Work}

In this paper we show results for the $\Lambda$CDM choice of the
initial fluctuation spectrum for which ($\Omega_0$, h, $\sigma_8$)
is (0.3, 0.7, 0.9), and $\Lambda$ = 1 - $\Omega_0$. Here $\Omega_0$
is the density in units of the critical density today, the Hubble
constant today is $H_0$ = 100$h$km s$^{-1}$ Mpc$^{-1}$, and
$\sigma_8$ describes the amplitude of the initial fluctuation
spectrum. In this work we utilized a simulated dataset from the
Virgo Consortium (the GIF simulation) that consisted of $256^3$
particles in a box of size $L = 141$Mpc/$h$. The correlation function
was calculated using the Landy-Szalay estimator and the NPT code.

Implementing the environmental dependence into the halo mass
function, we found that there was an observable upward shift of
the correlation function for haloes in dense regions with respect
to the correlation function for haloes in all regions. This is
shown in the left panel of Fig. 1.
\begin{figure}
\centering
\begin{minipage}[t]{6.6cm}
\includegraphics[width=1.\textwidth]{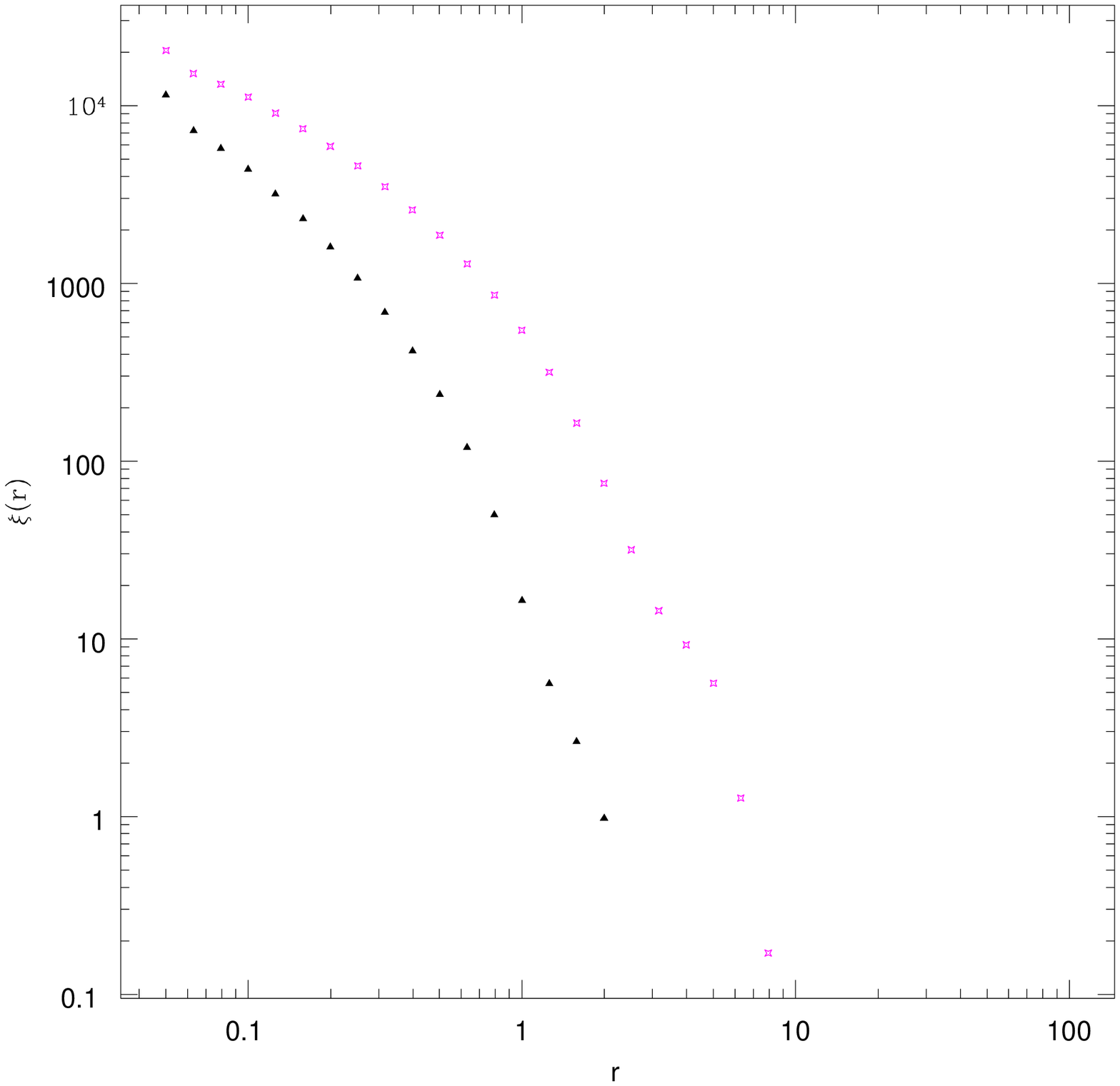}
\end{minipage}
\hskip 0.1cm
\begin{minipage}[t]{6.6cm}
\includegraphics[width=1.\textwidth]{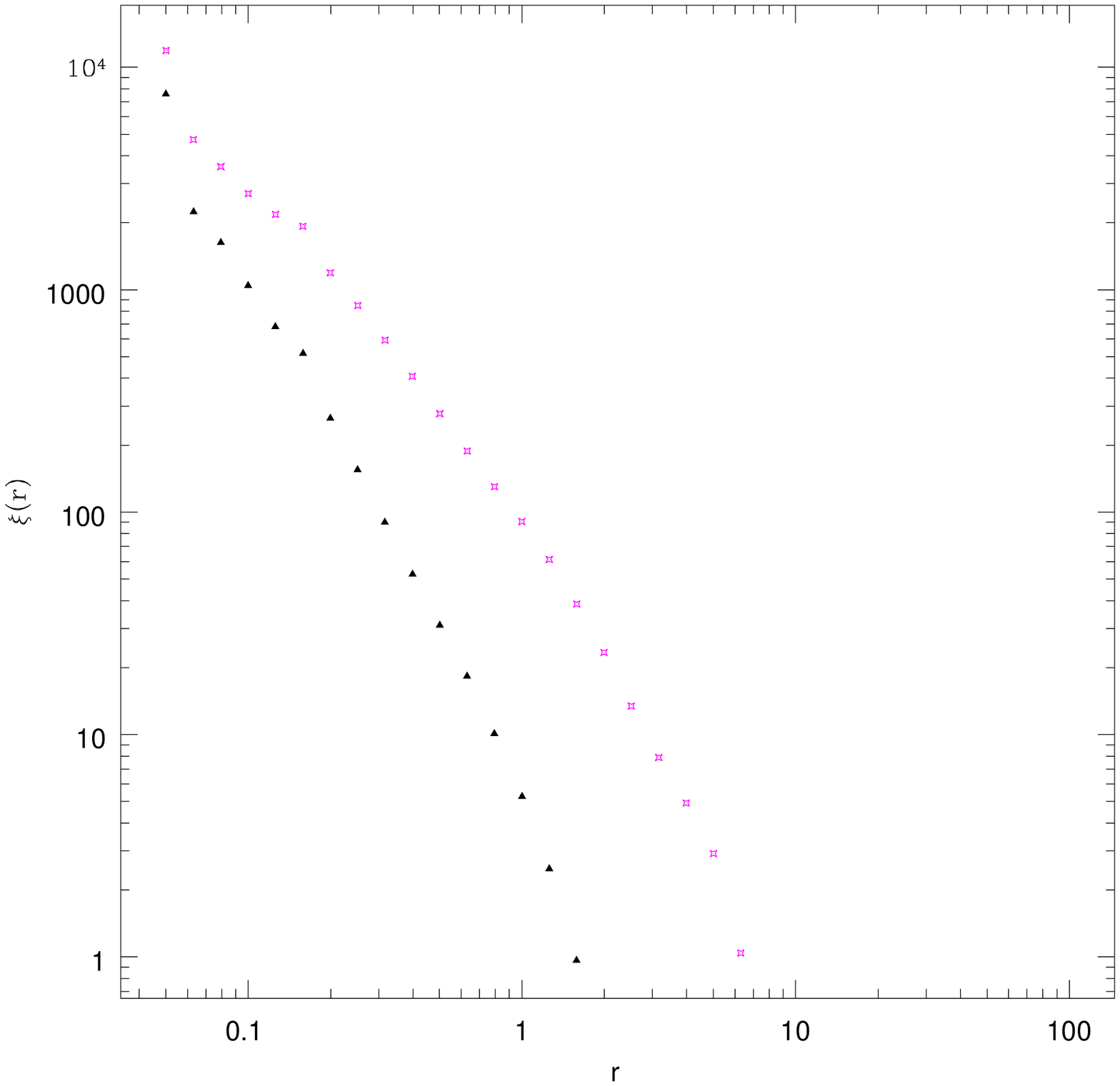}
\end{minipage}
\caption{Left panel: correlation function for dark matter in the
$\Lambda$CDM GIF simulations. The upper curve is the contribution
from particles in the densest regions (defined on a radius of 5
Mpc/h) and the lower curve is from particles in underdense
regions. Right panel: the same, but for galaxies}
\end{figure}

This model was extended to correlation functions describing
galaxies by implementing the halo occupation numbers. And once
again we measured an observable upward shift for galaxies in
haloes that were in denser regions (right panel of Fig 1). 
The process of ongoing
work is to derive a simple summary of simulation results in terms
of empirical analytical expressions describing the environmental
dependence in terms of specific parameters (ie. the halo mass).

With data from the Sloan Digital Sky Survey (SDSS) the models can
be confronted with statistical data of unprecedented quality to
obtain stringent constraints on the halo occupation numbers, and
therewith on both cosmological parameters and galaxy formation
models. This will be the focus of future work.

We will take this approach further by considering the conditional
luminosity function (CLF). The advantage of the CLF over the halo
occupation function $N_g(m)$ is that it will allow us to address
the clustering properties of galaxies as a function of luminosity.
In addition, the CLF is a direct link between the halo mass
function and the galaxy luminosity function. Therefore, the CLF is
not only constrained by the clustering properties of galaxies, as
in the case with $N_g(m)$, but also by the observed galaxy
luminosity function.

\begin{acknowledgments}
We would like to thank the Virgo consortium for kindly making the
simulations available, and the Pittsburgh Computational
Astrostatistics Group (PiCA) for the NPT code.
\end{acknowledgments}

\end{document}